\documentclass[apj,onecolumn]{emulateapj}   
\usepackage{graphicx} 
    \usepackage{epstopdf} 
    \usepackage{longtable} 
    \usepackage{natbib}
    \usepackage{chngpage}

\begin{document}

\title{Transit Analysis Package (TAP and autoKep):\\IDL Graphical User Interfaces for Extrasolar Planet Transit Photometry}
\author{
J. Zachary Gazak\altaffilmark{1},
John A. Johnson\altaffilmark{2,3},
John Tonry\altaffilmark{1}, 
Jason Eastman\altaffilmark{4}, 
Andrew W. Mann\altaffilmark{1},
Eric Agol\altaffilmark{5}
}

\begin{abstract}We present an IDL graphical user interface-driven software package 
designed for the analysis of extrasolar planet transit light curves.   The Transit Analysis Package (TAP) software uses Markov Chain Monte Carlo (MCMC) techniques to fit light curves using the analytic model of  \cite{2002ApJ...580L.171M}.  The package incorporates a wavelet based likelihood function developed by \cite{2009ApJ...704...51C} which allows the MCMC to assess parameter uncertainties more robustly than classic $\chi^2$ methods by parameterizing uncorrelated ``white'' and correlated ``red'' noise.  The software is able to simultaneously analyze multiple transits observed in different conditions (instrument, filter, weather, etc).  The graphical interface allows for the simple execution and interpretation of Bayesian MCMC analysis tailored to a user's specific data set and has been thoroughly tested on ground-based and Kepler photometry.  AutoKep provides a similar GUI for the preparation of Kepler MAST archive data for analysis by TAP or any other analysis software.  This paper describes the software release and provides instructions for its use. 

\end{abstract}

\email{zgazak@ifa.hawaii.edu}

\altaffiltext{1}{Institute for Astronomy, University of
Hawai'i, 2680 Woodlawn Dr, Honolulu, HI 96822}

\altaffiltext{2}{Department of Astrophysics, California Institute of Technology, MC 249-17, Pasadena, CA 91125}
\altaffiltext{3}{NASA Exoplanet Science Institute (NExScI)}
\altaffiltext{4}{The Ohio State University, Columbus, OH 43210}
\altaffiltext{5}{Department of Astronomy, Box 351580, University of Washington, Seattle, WA 98195}
\maketitle

\section{Initial Code Release}

TAP and autoKep are available for download at \begin{tt}http://ifa.hawaii.edu/users/zgazak/IfA/TAP.html \end{tt} along with descriptions and example videos.  A full paper will be made available shortly in which we describe comparison tests with published ground-based and Kepler
light curves, as well as new data.  A preprint is available at the TAP website. 

In \S\ref{TAP} we provide a manual for the use of the MCMC fitting feature of TAP, and \S\ref{autoKep} provides a manual for autoKep, which is designed to streamline the reduction of pipeline-reduced MAST Kepler light curves into a form usable by TAP or any transit analysis code.

\section{Attribution}
If you use TAP in a publication, please cite this paper and the following, which provide direct contributions to the code and/or methods used by TAP: 

\begin{enumerate}
\item \citealt{2002ApJ...580L.171M} for the Mandel \& Agol transit light curve model.
\item \citealt{2009ApJ...704...51C}, for the wavelet-based red noise treatment
\item \citealt{eastman2011}, for EXOFAST implementation of the Mandel \& Agol model

\end{enumerate}


\clearpage

\section{TAP Manual:}
TAP is designed to input transit light curves and fit them using the Mandel \& Agol model.  The input light curve data must be in ascii files in which the first column contains time in Julian Days (HJD) and the second column contains flux (normalized to an ``out of transit'' value of unity).  Any extra columns are ignored by TAP.

\label{TAP}
\begin{enumerate}
\item Starting TAP:
\begin{description}
\item[a) Start an IDL session]
\item[b) Run `tap' at the idl command line] This creates an instance of the TAP GUI.  No further interaction with the IDL command line is necessary.
\end{description}
\item Load one or multiple transits:
\begin{description}
\item[a) Click on ``Transit File"] under the ``Load Transit" menu
\item[b) Select your file, click ``ok" in the pop-up window]  Files must be in ascii format.  TAP uses the first two columns.  First column must be a timestamp in days.  Second column must be flux and should be normalized to an out of transit value of 1.  Additional columns are not used. 
\item[c) Set Resampling mode]  For long integration observations (such as Kepler's 29.4244 minutes per datapoint), the Mandel and Agol light curve must be calculated on a finer resampled time scale and then rebinned to the observed cadence.  If this is the case, select the `Rebin to ``Input Integration''' option.  In this case, the Input Integration box will become active.  Set your integration time and the number of samples (per datapoint) for resampling.  In general, N=4 samples is sufficient for Kepler light curves.  See  \cite{2010arXiv1004.3741K} for a more complete description.  
\item[d) Click ``Load Transit"]  The plot window will display the transit light curve and residuals in the bottom plot.  The initial transit guess will appear as a blue curve, and the residuals
	   will be plotted as distance from that blue curve.  The default strength
	   for "correlated red noise" is 0 (i.e. no red noise).  If you choose to
	   fit for red noise as well, a red curve will also be overplotted, showing
	   instead the transit model with the red noise signal added to it.
	   
\item[NOTE] To load multiple transits, either repeat the above steps with each ascii file or prepare an ascii file with all transits separated by a line of -1's matching the number of columns in your file.  TAP will automatically load each transit separately.  
\item[Restore an existing setup] If you have started an MCMC calculation and want to restore that exact setup, click on the ``Setup File" button under the ``Load Transit" menu.  Select a file, which TAP automatically creates, called ``TAP\_setup.idlsav" which is automatically saved into the directory of an MCMC run.  The setup will be restored, and changes can then be made before beginning a new MCMC calculation.
\end{description}

\item Adjust parameters for multiple transits:
\begin{description}
\item[a) Click on the ``Manage Transits" menu button]
\item[b) General items] Sliders which control which transit is ``active" and options for plot scaling.  The ``active transit" is the one which is affected by manual changes to the fit parameters.  It is labeled with blue arrows in the plot and its current parameters are plotted in the Parameter box to the left of the plot.	
\item[c) Setup Inter-Transit Links] This button launches a new window which contains an entry for each parameter of each transit.  The values in the boxes represent the current ``set'' the parameter belongs to.  Parameters in any row with the same set are locked together, such that those transits have the same parameter values and must evolve together in the MCMC calculations.  
\end{description}

\item Fit the transit light curve:
\begin{description}
\item[a) Click on the ``Fit''  menu button]
\item[b) ``Manual Parameter Adjustment"] this button opens a new window with 
		 13 sliders.  These represent the possible free parameters for MCMC 
		 fitting.  The values can be adjusted by dragging the sliders or by
		 typing in values and pressing "return".  The light curve plotted
		 in the main window will update automatically.

		   \textbf{``System Parameters"} The parameters of the transiting system from
		       the Mandel \& Agol (2002) analytic light curve.  

		     \textbf{``Quadratic Limb Darkening"} Linear and quadratic Limb Darkening
		       Coefficients.  

		     \textbf{``Data Specific Parameters"} A linear correction for airmass or 
		       normalization.  Slope and Y-intercept of the "out of transit"
		       level to account for any trend in the data.  Also Noise parameters: Strength of "White" uncorrelated 
		       gaussian noise and "Red" (1/f) correlated noise.  

\item[c) ``Parameter Limits and Locks"] this button opens a new window with one 
	         row for each of the 13 possible free parameters. 
		     
		    \textbf{``Min/Max":} The minimum and maximum of the sliders for "Manual 
		        Parameter Adjustment", and, if "Apply Limits to Fitting" is 
			selected, the MCMC is not allowed to wander beyond these
			limits.   
		     
		     \textbf{``Lock":} Select this to lock a parameter during fitting.  This is
		        used for parameters that are unable to be determined from the
			transit light curve, such as period in the case of a single
			light curve.
		   
		       \textbf{``MCMC Accept Rate":} This is the percent of jumps for which the MCMC
		        algorithm will optimize the jump betas.  The typical value 
			for efficient MCMC convergence is ~44
			default value.
		    
		        \textbf{``Apply Limits to Fitting":} Will force the MCMC execution to explore
		        a parameter space limited by the "Max/Min" values.
         
\item[d) ``MCMC Parameters''] Set the number of chains and links per chain for the MCMC execution.
\item[e) ``Gaussian Priors"]  Penalty terms allowing a parameter to vary in
		         the MCMC analysis but under a gaussian penalty term.  The term
			 works as a modification to the Likelihood of any particular fit
			 by adding a term of the form exp$[-(\textbf{value}-x_i)^2/\textbf{sigma}^2]$, where value and sigma correspond to the set priors and $x_i$ is the current MCMC value of that parameter.
			 This is useful when a parameter is theoretically known but should
			 be allowed to wander, most commonly the limb darkening parameters.
			 The code will then find an ideal solution between the ``Value" and
			 mathematically most likely solution.  Set a value, sigma, and click
			 the ``Enable" box to activate this feature.
\item[f) ``Execute Chain"] This will open a new window and freeze the base window.  The
	           new window will display the progress of the MCMC chain analysis and the
		   plot window on the frozen TAP base will cycle through the evolving 
		   parameter distributions.  The GUI will create a new directory to contain the MCMC run in the directory from which TAP was started.

\end{description}

\item Analyze MCMC Chain(s)

\begin{description}
\item[a) Click on the ``MCMC Inference''  menu button]
\item[b) Select options for output] A .tex document is automatically created containing details of the MCMC run and tables of results.  Beware of creating ascii files of the MCMC chains--they can become quite large.  Each output ascii file contains in its header a code snippet to load into IDL using 'readcol' 
\item[c) Click the ``MCMC Save File" Button] A dialog will appear to allow you to select a `TAP\_setup.idlsav' file.  Select this file in any MCMC directory which has completed at least 1 MCMC chain and press ``ok''.
\item[d) Click the ``Load" button] TAP does the rest,  automatically loading, combining, and analyzing the completed chains. You will find the output files in the root MCMC directory.  You can perform the MCMC inference as often as you would like, and as new chains complete the software will automatically utilize the full available results.
\end{description}

\end{enumerate}

\begin{figure*}[ht]
\centering
\includegraphics[width=7in]{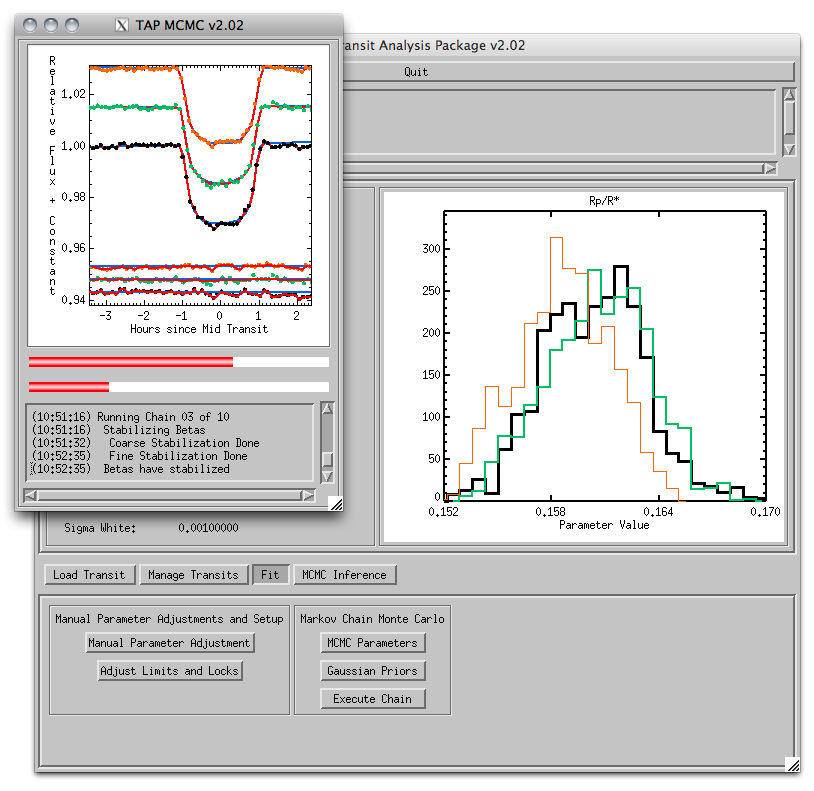}
\caption{The Transit Analysis Package (TAP) Graphical User Interface while executing a Markov Chain Monte Carlo analysis on three light curves of WASP-10b.  The main background widget freezes during MCMC calculation as a new smaller widget is spawned in the foreground.  The MCMC widget provides updates on the status of the analysis process while plotting every tenth $x_n$ model state with a blue analytic model and red curve representing the model and red noise.  The top and bottom progress bar represent the current chain and overall completeness, respectively.  During MCMC execution, the window in the main widget cycles through the currently explored probability density distributions for different parameters so that a user may monitor the progress of the analysis. }.  
\label{fig:tap1}
\end{figure*}

\clearpage

\section{autoKep Manual:}
autoKep is designed to load in *llc.fits (pipeline reduced light curve) files from the Kepler MAST archive.  The transits are output as ascii files and (optionally) loaded into TAP.

\label{autoKep}

\begin{enumerate}
\item Starting autoKep:
\begin{description}
\item[a) Start an IDL session]
\item[b) Run `autokep' at the idl command line] This creates an instance of the autoKep GUI.  No further interaction with the IDL command line is necessary.
\end{description}
\item Load one or multiple Kepler MAST .fits files:
\begin{description}
\item[a) Click on ``Fits File" under the ``Load Fits" menu] in the pop up dialog, select a Kepler MAST .fits file, then press ``ok''
\item[b) Click ``Load Fits File'' or ``Load all .fits''] The former option will load the currently selected fits, the latter loads all .fits files in the directory from which autoKep was started.
\end{description}
\item Detect Transits
\begin{description}
\item[a) Click on ``Detect Transits" menu item]  The GUI workspace will change.
\item[b) Find the first transit]  Either click on ``Auto Detect'' and give the software a shot at automatically finding the transit or click ``Manual Detect", then click to the left and right of a transit in the top (white backgrounded) plot.  In either case, if autoKep finds a plot a popup window will ask you to verify the detection.  Do so.
\item[c) Find the second transit] Repeat the procedure by clicking a button in the ``Second Transit'' submenu.  You should select an epoch immediately neighboring the initial transit detection, as autoKep will then use these two transits to estimate orbital period and ``map'' the remaining transits in yellow. 
\item[d) Map the transits] If the yellow regions do not cover the transits in the top plots, click on the ``Adj. Yellow Maps'' button.  A widget will open which allows you to modify period and duration (and thus the location and witdth, respectively, of the predicted transit epochs) anchored to the \textbf{second} detected transit.  Once the yellow maps look OK, click on the ``Map Transits'' button.  This button searches each yellow "epoch" for a transit signature and stores each as a transit light curve.
\end{description}
\item Adjust the Transits
\begin{description}
\item[a) Click on ``Adjust Transits" menu item]  The GUI workspace will change.
\item[b) Adjust the Duration of the Transits] The left two horizontal sliders adjust the duration of the transit.  The top slider is the transit event itself (red points).  These points are ignored when out of transit variations are fit, so it is important to make sure the entire transit is covered.   The bottom slider marks the extent of the out of transit light curve (blue points) which will be fit with a polynomial trend.  Enough points are needed to properly fit the curve.  All transits are fit with the same duration and OOT windows.
\item[c) Active transit] the vertical slider sets the active transit, which is plotted in the lower left plot and in color in the lower right plot.  This slider is useful for assuring that the out of transit trend fits each transit well.  
\item[d) Fit OOT Trends] the rightmost pair of horizontal sliders adjust the polynomial order of the OOT trend.  The top slider adjusts only the trend of the active transit, while the bottom slider adjusts all transits at once (to the same value).  
\end{description}
\item Export the Transits
\begin{description}
\item[a) Click on ``Export Transits" menu item]  The GUI workspace will change.
\item[b) Set the ``Write Region''] The extent (in days) of the exported transit.  To reduce MCMC calculation time it may be pertinent to cut down the output, being certain to leave sufficient out of transit points for the MCMC model fitting.  
\item[c) Export Options] A user may choose whether or not to apply the OOT corrections set in the ``Adjust Transits'' menu, and may also choose to automatically export the transit to TAP for analysis (the file is also written as a .ascii file).
\item[d) Root Filename] Type in the base filename, to which .ascii will be appended, for the output transit.  A single file is written, with 3 columns: BJD, normalized flux, and photometric error.  Each transit is separated by a line of -1's, as is the standard form for loading multiple transits into TAP from a single file.
\item[e) Export Transits] Click the lower Export Transits button.  If you have chosen the
``Export to TAP'' option, autoKep will automatically exit and TAP will start.  Otherwise, the GUI remains active for additional changes.
\end{description}

\end{enumerate}

\begin{figure*}[ht]
\centering
\includegraphics[width=5in]{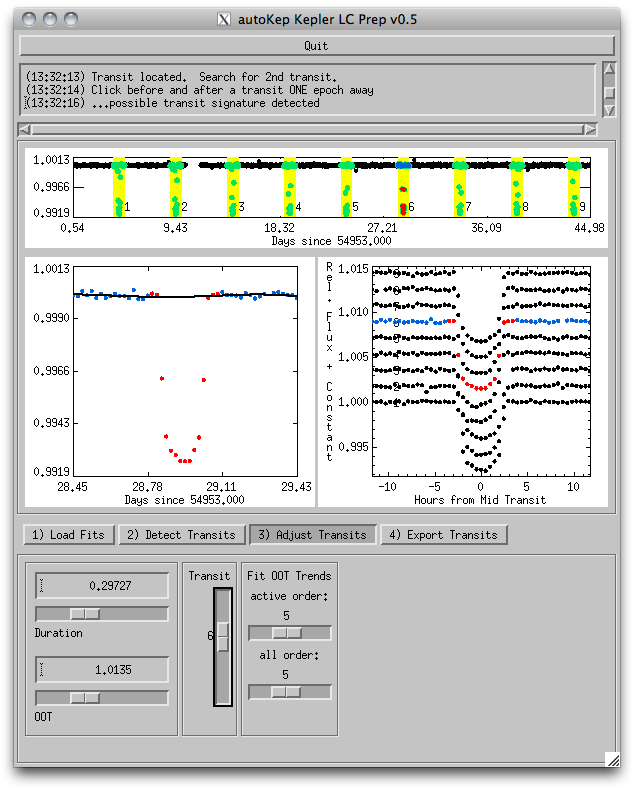}
\caption{The autoKep Graphical User Interface.}
\label{fig:autokep}
\end{figure*}

\clearpage

\bibliography{}

\begin{thebibliography}{4}
\expandafter\ifx\csname natexlab\endcsname\relax\def\natexlab#1{#1}\fi
\expandafter\ifx\csname bibnamefont\endcsname\relax
  \def\bibnamefont#1{#1}\fi
\expandafter\ifx\csname bibfnamefont\endcsname\relax
  \def\bibfnamefont#1{#1}\fi
\expandafter\ifx\csname citenamefont\endcsname\relax
  \def\citenamefont#1{#1}\fi
\expandafter\ifx\csname url\endcsname\relax
  \def\url#1{\texttt{#1}}\fi
\expandafter\ifx\csname urlprefix\endcsname\relax\def\urlprefix{URL }\fi
\providecommand{\bibinfo}[2]{#2}
\providecommand{\eprint}[2][]{\url{#2}}

\bibitem[{\citenamefont{{Mandel} and {Agol}}(2002)}]{2002ApJ...580L.171M}
\bibinfo{author}{\bibfnamefont{K.}~\bibnamefont{{Mandel}}} \bibnamefont{and}
  \bibinfo{author}{\bibfnamefont{E.}~\bibnamefont{{Agol}}},
  \bibinfo{journal}{\apjl} \textbf{\bibinfo{volume}{580}},
  \bibinfo{pages}{L171} (\bibinfo{year}{2002}).

\bibitem[{\citenamefont{{Carter} and {Winn}}(2009)}]{2009ApJ...704...51C}
\bibinfo{author}{\bibfnamefont{J.~A.} \bibnamefont{{Carter}}} \bibnamefont{and}
  \bibinfo{author}{\bibfnamefont{J.~N.} \bibnamefont{{Winn}}},
  \bibinfo{journal}{\apj} \textbf{\bibinfo{volume}{704}}, \bibinfo{pages}{51}
  (\bibinfo{year}{2009}).

\bibitem[{\citenamefont{{Eastman} et~al.}(2011)\citenamefont{{Eastman}, {Agol},
  and S.}}]{eastman2011}
\bibinfo{author}{\bibfnamefont{J.}~\bibnamefont{{Eastman}}},
  \bibinfo{author}{\bibfnamefont{E.}~\bibnamefont{{Agol}}}, \bibnamefont{and}
  \bibinfo{author}{\bibfnamefont{G.}~\bibnamefont{S.}}, \bibinfo{journal}{in
  prep}  (\bibinfo{year}{2011}).

\bibitem[{\citenamefont{{Kipping}}(2010)}]{2010arXiv1004.3741K}
\bibinfo{author}{\bibfnamefont{D.~M.} \bibnamefont{{Kipping}}}
  (\bibinfo{year}{2010}), \eprint{arxiv: 1004.3741}.

\end{thebibliography}

\end{document}